\documentclass[aps,prl,reprint,superscriptaddress,floatfix,amsmath,showpacs]{revtex4-1}

\usepackage{graphicx}
\usepackage{textcomp}
\usepackage{ulem} 

\begin{document}
\title{Radiation of caustic beams from a collapsing bullet}

\author{M. P. Kostylev}
\affiliation{School of Physics, M013, University of Western Australia, Crawley, WA 6009, Australia}

\author{A. A. Serga}
\affiliation{Fachbereich Physik and Forschungszentrum OPTIMAS, Technische Universit\"{a}t
Kaiserslautern, 67663 Kaiserslautern, Germany}

\author{B. Hillebrands}
\affiliation{Fachbereich Physik and Forschungszentrum OPTIMAS, Technische Universit\"{a}t
Kaiserslautern, 67663 Kaiserslautern, Germany}

\begin{abstract}
Collapse of an intense (2+1)-dimensional wave packet in a medium with cubic nonlinearity and a two-dimensional dispersion of an order higher than parabolic is studied both theoretically and experimentally. The carrier waves are microwave backward volume spin waves which propagate in a stripe made from a thin ferrimagnetic film and the packet is a spin-wave bullet. We show that before being self-destroyed the bullet irradiates untrapped dispersive waves, which is in agreement with a previous theoretical prediction.
Since, in addition, the ferromagnetic medium is characterized by an induced uniaxial anisotropy, this radiation takes the form of narrow beams of continuous waves at very specific angles to its propagation direction. Based on our theoretical calculations we find that these beams are caustic beams and the angles are the characteristic spin-wave caustic angles modified by the motion of the source. 	
\end{abstract}

\pacs{75.30.Ds, 05.45.Yv, 42.25.Fx}



\maketitle Solitons while perturbed are able to emit small-amplitude radiation with frequency detuned far from the soliton frequency \cite{Akmediev_and_Carson, KarpmanPRB47,Wai_Chen_Lee}. This ability, generally recognized as striking evidence of soliton's wave nature \cite{Skryabin_and_Gorbach}, has gained dramatically in importance in recent years with the discovery of super\-continuum generation in photonic crystal fibres whose applications for frequency comb generation in metrology, spectroscopy and imaging are more than just impressive (see e.g. \cite{Dudley_and_Taylor}). It was shown that dispersion radiation from solitons is responsible for a major spectral part of the super\-continuum (see e.g. \cite{Agrawal, Skryabin_and_Gorbach}).

One has to note that small-amplitude radiation from solitons is possible, provided the dispersion relation for the medium deviates from the parabolic law $\omega(k)=\omega_0+v(k-k_0)+D/2(k-k_0)^2$ built-in to the standard (i.e. parabolic) Cubic Nonlinear Schr\"{o}dinger Equation (CNSE) which governs envelope solitons in optical fibers \cite{Hasegawa}, deep-water solitons \cite{Zakharov}, and spin-wave envelope solitons in magnetic films \cite{Zvezdin}:
\begin{equation}
\frac{\partial a}{\partial t} + i\omega_0 a + v\frac{\partial a}{\partial z} - i\frac{D}{2}\frac{\partial^2 a}{\partial z^2}+iN |a|^2a=0
\nonumber
\end{equation}
Here $a$ is the envelope function for the propagating wave packet, $z$ is the direction of propagation,
and $N$ is the nonlinear coefficient (the term $iN|a|^2a$ is called ``cubic nonlinearity'' and gives rise to the equation name).

The dispersion law is expressed as a series expansion of the frequency $\omega$ in terms of the wave number $k$ of plane linear waves supported by the medium. The sign of the second-order dispersion coefficient $D=\partial^2 \omega(k)/\partial k^2$ is of paramount importance for soliton formation: solitons are formed provided $DN<0$ \cite{higher-order dispersion}.

Karpman \cite{KarpmanPLA160} theoretically predicted that in the same CNSE framework the fourth-order two-dimensional (2D) dispersion $\omega(k_y,k_z)$ with a set of coefficients of proper signs should lead to wave irradiation from (2+1)-dimensional wave packets $a(x,y,t)$ as well. He considered an axially symmetric wave packet in an isotropic medium. In his work and in subsequent numerical simulations \cite{Karpman_and_Shagalov} it was shown that a localized (2+1)D wave packet is destroyed by radiating energy into a non-trapped circular wave.

To the best of our knowledge this radiation has not been observed so far. Interestingly, a similar effect of conical Cerenkov radiation from optical bullets in an axially symmetrical 3D geometry \cite{Nibbering} has attracted a lot of attention \cite{Gaeta}. The conical radiation does not originate from a dispersion of a higher-order but from a higher-order nonlinearity. Thus, it can be characterized as an effect ``dual'' to one Karpman predicted.

In this work we experimentally demonstrate the non-trapped radiation from a collapsing bullet originating from the higher-order dispersion. This phenomenon is observed for a spin-wave bullet propagating in a thin magnetic film. Furthermore, we observe two  extra features which were not included in the original theory \cite{KarpmanPLA160}. First, (i) we find that in a medium with uniaxial anisotropy this radiation takes the form of two narrow-aperture beams. Second, (ii) as the medium is characterized by a dispersion of backward type, the radiation is in the backward direction with respect to the bullet propagation direction. This is in contrast to Cerenkov conical radiation which is forward \cite{Nibbering}.

Stable (2+1)D localized nonlinear spin-wave excitations, termed spin-wave bullets, have been previously observed in thin ferrimagnetic films of yttrium iron-garnet (YIG) magnetized along the propagation direction
\cite{Linear_nonlinear_diffraction_dipolar_spin_waves_yttrium_iron_garnet_films_observed_space-time-resolved_Brillouin_light_scattering,
Self-Generation_Two-Dimensional_Spin-Wave_Bullets,
Parametric_Generation_Forward_Phase-Conjugated_Spin-Wave_Bullets_Magnetic_Films, PRL-guidedBullets2008}. The waves propagating along the field are called backward volume magnetostatic spin waves. They are waves of backward nature, which means that their group velocity and wave vector point in opposite directions.

Two-dimensional pulses are intrinsically unstable and undergo nonlinear narrowing leading to collapse as nonlinearity overcompensates linear broadening due to 2D parabolic dispersion. Weak magnetic losses in a real magnetic film may balance the nonlinear narrowing. This results in a quasi-2D spatially localized bell-shaped waveform called a bullet and ensures its stability for some distance of propagation
\cite{Linear_nonlinear_diffraction_dipolar_spin_waves_yttrium_iron_garnet_films_observed_space-time-resolved_Brillouin_light_scattering}.
This regime is well described by the (2+1)D parabolic CNSE. The stable bullets are observed in a certain range of initial powers for the wave packets. For larger input powers waveform collapse is unavoidable
\cite{Linear_nonlinear_diffraction_dipolar_spin_waves_yttrium_iron_garnet_films_observed_space-time-resolved_Brillouin_light_scattering}.
The advanced stage of the collapse is beyond the limits of applicability of the parabolic CNSE, and the specific
scenario for destruction of the collapsing wave packet (``collapse scenario'') should be governed by processes of
higher-order.

\begin{figure}
    \scalebox{1}{\includegraphics{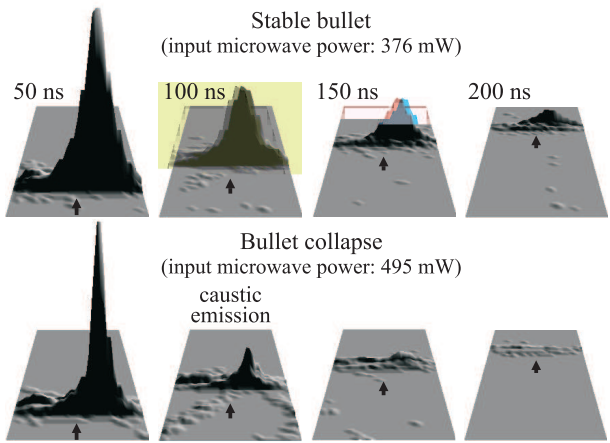}}
    \vspace*{-0.0cm}\caption{Snapshots of spin-wave packets measured at different input powers and at 50~ns time intervals.
    Arrows show the bullet propagation direction. It coincides with the direction of the bias magnetic field $\textbf{H}$
    and the coordinate axis $\textbf{z}$.}
\end{figure}

Our experiment is carried out using a longitudinally magnetized yttrium iron garnet (YIG) film stripe which is 2.5~mm wide ($w=2.5$~mm) and $5~\text{\textmu m}$ thick. The magnetizing field $H$ is 1831~Oe. The spin waves are excited by a rf magnetic field created with a 25~$\text{\textmu m}$ wide microstrip antenna placed across the stripe and driven by 20~ns long microwave current pulses at a carrier frequency of 7.125~GHz. The spatio-temporal behavior of the traveling spin-wave packets is probed by means of space- and time-resolved Brillouin light scattering spectroscopy
\cite{Physics_Reports}. For relatively small input powers we observe formation of quasi-stable (2+1)D wave packets -- guided spin-wave bullets (Fig.~1, upper row) -- reported previously for the same geometry \cite{PRL-guidedBullets2008}. If we increase the power beyond the range of bullet stability the wave packet collapses (Fig.~1, lower row). The most prominent feature of the collapse is the pair of rays irradiated from the packet in the backward direction (second panel in the lower row of Fig.~1). The left panel in Fig.~2 gives a detailed view of the beam emission. One clearly
sees that the rays have narrow apertures and are directed at well defined angles to the longitudinal axis of the ferromagnetic stripe: the value of the angle between the rays is $64^\circ$.
\begin{figure}
    \scalebox{1}{\includegraphics{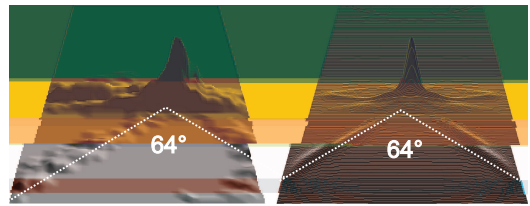}}
    \vspace*{-0.0cm}\caption{Irradiation of caustic beams from the collapsing bullet.
   Left panel: experiment. Right panel: simulation.}
\end{figure}

In order to reveal the origins of the irradiation we carry out numerical simulations. An original reciprocal-space approach is used which effectively accounts for all dispersion orders of the real spin-wave dispersion \cite{KostylevPRB76_224414, PRL-guidedBullets2008}. In the reciprocal space CNSE takes the form of a system of coupled equations for the amplitudes of the wave packet's spatial Fourier components:
\begin{equation}
\begin{split}
i\frac{\partial F_{n,k}(a)}{\partial t}+(\omega_{n,k}+i\Gamma-\omega_0)F_{n,k}(a) \nonumber
+ NF_{n,k}(|a|^2 a) \\ = f_{n,k}(t),
\end{split}
\end{equation}
where $a=a(y,z,t)$ is the spin-wave scalar amplitude and
\begin{equation}
\begin{split}
F_{n,k}(a)=\frac{1}{2\pi w}\int_{-w/2}^{w/2}\sin\left(\frac{n\pi y}{w}\right)dy
\int_{-\infty}^{\infty} a e^{-ik_zz} dz \nonumber
\end{split}
\end{equation}
denotes a 2D Fourier transform which is continuous along the stripe ($k_z$ takes continuous values in the model) and discrete in the transverse direction ($k_y = n\pi/w, n=1,2,3,\ldots$). This term is responsible for nonlinear coupling of Fourier components with different values of $n$ and $k_z$. $\Gamma$ denotes the coefficient of spin-wave linear damping for the medium and $f_{n,k}(t)$ is the Fourier transform of the time envelope of the driving microwave field which excites the spin-wave packet at the entrance into the waveguide medium. Similar to the experiment, $f_{n,k}(t)$ is
taken in the form of a 20~ns-long rectangular pulse. The carrier frequency of this driving pulse is $\omega_0/2\pi$ (see Fig.~3).	

The term $\omega_{n,k} = \omega(k_y, k_z)$ is the 2D spectrum of spin waves, shown in Fig.~3 as $k_z$-dependence of frequency for a family of different discrete values of $k_y$ which we take here equal to $n\pi/w$. The calculated dispersion corresponds to the conditions of our experiment. Physically this $(n,k_z)$-representation describes a family of guided modes for a film waveguide. The transverse ($y$-)profiles of the width modes are given by $\sin(n\pi y/w)$ with $n$ indicating each particular mode. One sees that the dispersion slope $\partial \omega_{n,k} / \partial |k_z|$ is negative for all the modes. This reflects the fact that the modes are of backward type. Consequently, a localized wave packet which propagates in the positive direction of the axis $z$ has a carrier wave number $k_z<0$.

As previously shown for both 1D \cite{KarpmanPRB47} and isotropic 2D media \cite{KarpmanPLA160}, to ensure radiation from nonlinear waveforms it is important to have a dispersion law for a medium which contains terms higher than parabolic. More precisely, the curvature $\partial^2 \omega/\partial k^2$ should change sign along the dispersion curve. From Fig.~3 one sees that on one hand for each particular width mode the sign of $\partial^2 \omega_{n,k} / \partial k_z^2$ changes from negative to positive in the vicinity of $k_z=0$. On the other hand, the spectrum in Fig.~3 is quite dense. Considering this spectrum as quasi-continuous one retrieves the 2D continuous dispersion of plane spin waves in ferromagnetic films \cite{DamonEshbach} which is characterized by the variation of dispersion sign $\partial^2\omega/\partial k^2$ as a function of the propagation angle with respect to the direction of the applied field $\textbf{H}$ ($k^2= k_y^2+ k_z^2$, $\phi=\arctan(k_y/k_z)$).

Our simulations confirm the validity of Karpman's idea that the higher-order dispersion alone may give rise to a non-trapped radiation from a collapsing bullet. The result is shown in Fig.~2. (The same parameters as for Fig.~3 and standard for YIG films have been used in this calculation.)
One sees good agreement with the experiment. In particular, we obtain the same radiation angle of $64^\circ$ as in the experiment.

The remaining part of the paper is aimed to understanding the extra features (i) and (ii) (see above) of the observed radiation and to clarifying the specific radiation angle shown in Fig.~2.

Waves excited  by an immobile point source in the medium with the dispersion shown in Fig.~3 are prone to forming caustics \cite{PRL-caustics2010}.  It is appropriate here to give a short introduction to the phenomenon of caustics. In an anisotropic medium with a dispersion law $\omega(\textbf{k})$, the direction of the wave group velocity $\textbf{v}_g =\partial \omega(\textbf{k})/\partial \textbf{k}$ indicating the direction of energy flow does not, in general, coincide with the direction of the wave vector $\textbf{k}$. When the medium anisotropy is sufficiently strong, the direction of the group velocity of the waves in the vicinity of a certain wave vector $\textbf{k}_c$ may become practically independent of their wave vectors. In this case the energy of the waves is channeled along this specific direction, which is called \textit{caustic direction}, and forms a so-called \textit{caustic beam}. In our recent work \cite{PRL-caustics2010} we demonstrated this phenomenon for spin waves in a single crystal YIG film using a quasi-point excitation source whose diameter $d$ was smaller than $2\pi/k_c$. In this medium magnetocrystalline anisotropy is negligible; application of an external static field $\textbf{H}$ in the film plane imposes a uniaxial symmetry necessary for the formation of caustics. Two pairs of spin-wave caustic beams are directed at specific angles with $\textbf{H}$ \cite{Linear_nonlinear_diffraction_dipolar_spin_waves_yttrium_iron_garnet_films_observed_space-time-resolved_Brillouin_light_scattering,
Vashkovsky, CamleyPRB74, DemidovPRB80, PRL-caustics2010, CamleyPRB81}.

One notices that a collapsing spin-wave bullet in a magnetically saturated ferromagnetic film meets the size criterion  $d < 2\pi/k_c$ for a quasi-point source of linear excitation of caustic waves. Such a source located on the axis of the wave\-guide will excite all waveguide modes which have an anti-node on the axis. These are the symmetric modes $n=1,3,5, \ldots$ For an immobile linear source, the frequency of the excited waves is equal to the frequency of the source. This condition is indicated by the dashed horizontal line in Fig.~3: all the symmetric modes whose dispersion lines cross the dashed line will be excited with the frequency given by the ordinate of the point of the cross-section. It is a short exercise \cite{Vashkovsky} to find which of the modes are responsible for caustic formation: those modes are shown by the bold part of the dashed horizontal line. For the conditions of our current experiment the angle $\phi_I$ between the beams in each caustic pair should amount to $84^\circ$ as our calculation based on the theory in Ref.~\cite{CamleyPRB74} shows. Note that this value is significantly different from the angle seen in Fig.~2.

\begin{figure}
    \includegraphics{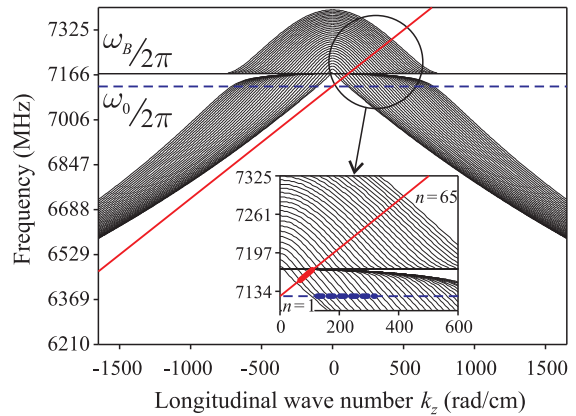}
    \vspace*{-0.0cm}\caption{Spectrum of the guided modes for the longitudinally magnetized film waveguide.
    The calculation assumes pinned spins at the stripe edges.
    The family of symmetric modes $n=1,3,...65$ is shown.
    The mode which is the lowest in frequency is the fundamental width mode for the wave\-guide $n=1$.
    Inset: a section of the spectrum close to the frequency $\omega_B /2 \pi$.
    The horizontal blue dashed lines in the main field of the figure and in the inset show the carrier frequency $\omega_0 /2 \pi$
    of the input microwave pulse.
    The oblique red lines show the Doppler-shifted frequencies of an excitation source which moves in the positive direction.
    The bold sections of these lines show the modes responsible for formation of the respective caustics.}
\end{figure}

This disagreement is caused by the motion of the bullet with the velocity $v = 2.5$~cm$/\text{\textmu s}$. The moving source is described by the red oblique line $\omega=\omega_0+vk_z$  which can be considered as the time-space Fourier transform of the moving point source (see e.g. Eq.~(14) in \cite{Haus}). The term $vk_z$ is obviously the Doppler frequency shift for the excited modes. One sees that now each mode is excited with its own frequency. Furthermore, one sees that crossings are possible only for positive $k_z$ values. Importantly, our calculations show that, similar to the previous case of the immobile source, the modes which satisfy the Doppler shift condition $\omega_{n,k}(k_z)=\omega_0+vk_z$ are also able to form caustic beams. The family of modes which are found to be responsible for formation of the modified caustics is shown by the bold section of the red oblique line.

The modified caustic angles are obtained by considering the family of slowness curves for the Doppler-shifted frequencies. A slowness curve is a constant-frequency line in the $(k_y,k_z)$ plane (Fig.~4a) calculated for a film which is continuous in both in-plane directions \cite{CamleyPRB74}. The group velocities of all plane waves which exist at the respective frequency are directed perpendicular to this curve. For this reason the direction of the energy flow (DEF) \cite{CamleyPRB74} is given by the normal to the slowness curve at the point where its curvature is zero. A range of frequencies for which the slowness directions are close to each other will contribute to formation of the modified caustics. The thin solid lines in Fig.~4 are the slowness directions for the frequencies which correspond to the crossing points of the oblique red line with the dispersion lines in Fig.~3. The bold dashed line in Fig.~4 runs across zeros of curvature for these curves. One sees that normals to the thin solid lines are collinear. This suggests that the normals define the DEF directions. The rigorous values for the angles between these directions and the bullet velocity are $\pm 136.5^\circ$ (Fig.~4b).
\begin{figure}
    \includegraphics{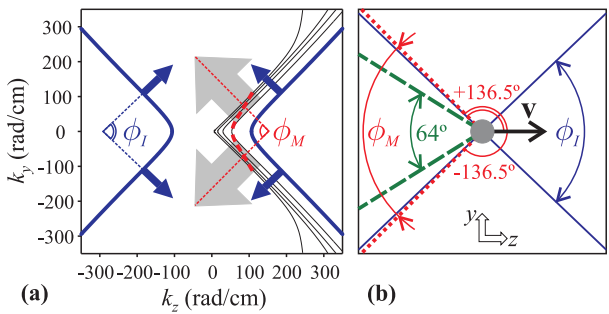}
    \vspace*{-0.0cm}\caption{Slowness curves for the immobile and the mobile excitation sources (a) and the respective angles (b).
   In (a): bold solid lines are the slowness curves for the frequency given by the horizontal dashed line in Fig.~3.
   Thin solid lines are the slowness curves for the Doppler-shifted frequencies.
   The red bold dashed line connects zero-curvature points of the thin solid lines.
   Dark blue arrows are normal to the thick solid line and show the caustic directions for excitation by an immobile source.
   Grey arrows are normal to the family of the thin solid lines and show the  energy flow directions (DEF) for the moving source.
   In (b): the arrow indicates the direction of source motion. Thin blue solid lines: caustic directions for the immobile source.
   Red dotted lines: two DEF for the moving source. Green dashed lines: directions along the observed caustic beams.}
\end{figure}

First, one sees that the angles exceed $90^\circ$ which reflects the fact that the carrier waves are the waves of backward nature. (Note that the Doppler frequency shift for backward waves is anomalous \cite{StancilPRB74,Chumak_PRB81}.) Second, one notices that the angle $\phi_M=87^\circ$ between the two DEF is increased by just $3^\circ$ with respect to $\phi_I=84^\circ$.

Furthermore, for a moving source the directions along the formed beams do not coincide with the directions of the energy flow.
It is instructive to define DEF via the angle of group velocity $\textbf{v}_g$. One finds $\phi_M=2 \arctan (|v_{gy}/v_{gz}|)$. The direction along a beam is related to the direction of $\textbf{v}_g$ by a Galilean transformation. As a result the angle between the two backwards-irradiated beams is given by $2 \arctan (|v_{gy}| /|v_{gz} - v|)$. Using this formula one obtains a value of $64^\circ$ which is in the excellent agreement with the results shown in Fig.~2.

In conclusion, we studied the collapse scenario for an intense two-dimensional wave packet propagating in a ferromagnetic medium and experimentally showed that before being self-destroyed the wave packet irradiates untrapped continuous waves. Our simulation proves that the observed effect is the fundamental phenomenon of wave emission from collapsing packets in materials with cubic nonlinearity and 2D-dispersion of an order higher than parabolic.
In addition, since the ferromagnetic-film medium used in the present study is also characterized by a uniaxial anisotropy, the observed radiation takes the form of narrow-aperture beams of continuous waves at very specific angles to bullet's propagation direction. These angles are larger than 90 degree, since the underlying wave dispersion is of backward type. This specific effect of radiation of untrapped waves along the caustic directions may exist for other 2D-media media as well, provided uniaxial anisotropy of dispersion is available or induced in the medium.

We acknowledge the financial support by Deutsche Forschungsgemeinschaft (TRR49) and the Australian Research Council. We also thank A.N.~Slavin and Yu.~Kivshar for stimulating discussions.


\begin{thebibliography}{24}
\bibitem{higher-order dispersion} We will call a dispersion relation containing term of orders higher than parabolic ``higher-order dispersion''.
\end{thebibliography}
\end{document}